\documentclass[%
preprint, 
 notitlepage,
 amsmath,amssymb,
 aps,
]{revtex4-1}

\usepackage{graphicx}
\usepackage{dcolumn}
\usepackage{bm}
\usepackage{color}
\usepackage{comment}
\usepackage[textsize=small]{todonotes}
\usepackage{placeins} 

\newcommand{\thcm}{\theta_\text{c.m.}}
\newcommand{\costhcm}{\cos \thcm}

\newcommand{\diffd}{\text{d}}
\newcommand{\diffcs}{\frac{\diffd \sigma}{\diffd \Omega}}

\newcommand{\norm}[1]{\ensuremath{\left| #1 \right|}}
\newcommand{\dist}[2]{\mathcal{D} \left[#1 , #2 \right]}
\newcommand{\B}{\mathcal{B}}
\newcommand{\T}{\mathcal{T}}
\newcommand{\R}{\mathcal{R}}

\newcommand{\M}{\mathcal{M}}

\newcommand{\fourv}{\ensuremath{\bm{\M}}}

\newcommand{\trans}[3]{\ensuremath{\bm{T}^{#1}_{#2,#3}}}
\newcommand{\polphoton}{\ensuremath{\B}}
\newcommand{\polnucleon}{\ensuremath{\T}}
\newcommand{\polhyperon}{\ensuremath{\R}}

\newcommand{\erroron}[1]{\ensuremath{\Delta #1}}

\newcommand{\loglike}{\ensuremath{\ln \mathcal{L}}}
\renewcommand{\Re}{\ensuremath{\operatorname{Re}}}

\usepackage[mathlines,columnwise]{lineno}

\begin{document}


\title{Model discrimination in pseudoscalar-meson photoproduction}

\author{J.~Nys}
\email{Jannes.Nys@UGent.be}

\author{J.~Ryckebusch}%
 
\affiliation{%
 Department of Physics and Astronomy, Ghent University, Belgium
}%

\author{D.\,G.~Ireland}
\author{D.\,I.~Glazier}

\affiliation{SUPA, School of Physics and Astronomy, University of Glasgow, United Kingdom}%

\date{\today}

\begin{abstract}

To learn about a physical system of interest, experimental results must be able to discriminate among models. We introduce a geometrical measure to quantify the distance between models for pseudoscalar-meson photoproduction in amplitude space. Experimental observables, with finite accuracy, map to probability distributions in amplitude space, and the characteristic width scale of such distributions needs to be smaller than the distance between models if the observable data are going to be useful. We therefore also introduce a method for evaluating probability distributions in amplitude space that arise as a result of one or more measurements, and show how one can use this to determine what further measurements are going to be necessary to be able to discriminate among models.
\end{abstract}

\maketitle


\section{Introduction}

\label{sec:introduction}
Nuclear and hadron physics have entered an era of high precision measurements from often very demanding experiments. In the planning stage, it is important to estimate the potential impact of a particular set of measurements. High impact experiments are ones in which there is a large potential for the data to constrain the models of the underlying physical processes of interest, typically by greatly reducing uncertainties in model parameters. An analysis of nucleon-nucleon scattering data, for example, with advanced statistical methods \cite{Perez:2014jsa} allows one to infer the parameters and corresponding errors in nucleon-nucleon potentials. Statistical methods that are designed to reliably infer parameters from experimental data are, however, not necessarily optimized to estimate the potential impact of various combinations of possible experiments. In other words, \emph{model discrimination} often requires different strategies than \emph{parameter estimation} within models \cite{MacKay:2002:ITI:971143, Wesolowski:2015fqa, Sangaline:2015isa}.

In this paper we lay out a framework that can be used to obtain estimates of the possible impact of (combinations) of polarization measurements in pseudoscalar-meson photoproduction from the nucleon (hereafter denoted as $\gamma N \rightarrow MB$). 
Information about the reaction amplitudes in a particular range of kinematics is the key to discriminating between two or more models. 
In imaging systems, the Rayleigh criterion is used to determine whether two or more light sources can be resolved from each other. We develop an analogue of this criterion which requires a measure of the distance between models in amplitude space, and a means of determining the characteristic spread of probability densities in amplitude space that result from measurement of observables.

Several models for the underlying reaction mechanisms of $\gamma N \rightarrow MB$ reactions are available. Some of the most common approaches are the coupled-channel (CC), isobar and hybrid isobar-Regge models. All of these aim to extract $s$-channel resonance content from experimental data. In most cases, model assumptions are required to describe other contributing mechanisms (referred to as ``the background''). After decades of research, however, the precise underlying resonance content is still under debate, and the list of known resonances changes with each edition of the Review of Particle Physics \cite{reviewofparticlephysics}.
A detailed knowledge of the reaction amplitudes as a function of kinematical variables should enable one to discriminate among various reaction models, but it is necessary to perform measurements of several $\gamma N \rightarrow MB$ polarization observables to access the reaction amplitudes. 

At fixed kinematics, four complex reaction amplitudes determine the $\gamma N \rightarrow MB$ dynamics. The kinematics are fixed by the invariant mass $W$ and the cosine of the center-of-mass (c.m.) scattering angle $\thcm$, and there is a one-to-one relation between $(W,\costhcm)$ and the Mandelstam variables $(s,t)$. It was suggested \cite{BDS1975,ChiangTabakin1997} that a selection of polarization measurements may lead to a situation where all reaction amplitudes are known to the extent that the outcome of any future experiment could be predicted. In Ref.~\cite{ChiangTabakin1997} it was shown that eight well-chosen observables suffice to unambiguously determine the amplitudes. One refers to a such a combination of observables as a ``complete set".  However, this is only true in a mathematical sense, and it has been established that there is no such thing as complete sets when dealing with data with finite error bars \cite{IncompletenessVrancx2013, NysIncompletenessComplete, Ireland, Wunderlich:2014xya, Tiator:2012ah}.

Two categories can be distinguished for polarization observables: single-polarization ($\mathcal{S} = \{\Sigma~(\textrm{beam}), T~(\textrm{target}), P~(\textrm{recoil})\}$) where only one of the initial and final state particles is polarized, and double-polarization that require two polarized particles. The latter category can be subdivided into three categories: beam-recoil ($\B\R = \{C_x, C_z, O_x, O_z\}$), beam-target ($\B\T = \{E,F,G,H\}$) and target-recoil ($\T\R = \{T_x,T_z,L_x,L_z\}$) observables \cite{Sandorfi:2010uv}. These are connected to the reaction amplitudes through bilinear relations (see e.g.\ Ref.\ \cite{IncompletenessVrancx2013}). We note that in practice, experiments are configured to have beam polarization, target polarization, the ability to determine recoil polarization or some combination thereof. Each of these experimental configurations are sensitive to different combinations of ``observables'', and so the observables are not usually measured in isolation~\cite{Dey:2010fb}.

Models that are fitted to the published observables, can in fact have very different reaction amplitudes. An example is the $\B\T$  double polarisation observable $E$ in $\vec{\gamma} \vec{p} \rightarrow \pi ^{+} n$ that was measured recently \cite{Strauch:2015zob}. Despite the availability of data for other observables,  the existing $\gamma p \rightarrow \pi ^{+} n$ models predicted a large range of values of $E$ at similar kinematic points (see Fig.~3 in Ref.~\cite{Strauch:2015zob}), pointing to substantial differences among the models at the amplitude level. The overall or ``global'' performance of two models can be compared by averaging  their least squared-distance to the measurements over all experimentally probed kinematics.  More restrictive is a ``local'' model discrimination, where models are compared at specific kinematics ($s,t$). A partial-wave analysis parameterizes the $\costhcm$ dependence of the reaction amplitudes at fixed $s$ and can be regarded as an analysis technique that falls in between ``local'' and``global''. In this work, we focus on the most local (and completely model-independent) form of amplitude analysis, but we note that in practice it is probable that model comparison will be done with partial wave analyses. The question that we aim to address is what kind of experimental results do we need to be able to discriminate between various models at specific kinematics. \\

In this work we use transversity amplitudes (TA), where particle spins are quantized in a transverse basis. The TA have so-called ``optimally simple" relations \cite{Goldstein:1974ym} to the observables, in which the single-polarization observables depend on the amplitude moduli only \cite{IncompletenessVrancx2013, NysIncompletenessComplete}. The transition amplitude $\trans{\polphoton}{\polnucleon}{\polhyperon}$ for a fixed photon $\polphoton$, nucleon $\polnucleon$ and baryon $\polhyperon$ polarization, reads
\begin{equation}
\trans{\polphoton}{\polnucleon}{\polhyperon} \equiv \overline{u}_B^{\polhyperon} \epsilon_{\polphoton}^\mu \hat{J}_\mu u_N^{\polnucleon} \,.
\end{equation}
The $u_B$ ($u_N$) denotes the recoil (target) Dirac spinor, $\hat{J}^\mu$ the interaction current and $\epsilon_{\polphoton}^\mu$ the $\gamma$-polarization four-vector.  For a linearly polarized photon along the $x$ or $y$ axis one has  $\epsilon_{\polphoton=x}^\mu = (0,1,0,0)$,  $\epsilon_{\polphoton=y}^\mu = (0,0,1,0)$.
The transversity basis is defined as
\begin{equation}
b_1 = \trans{y}{+y}{+y} \,,~
b_2 = \trans{y}{-y}{-y} \,,~
b_3 = \trans{x}{-y}{+y} \,,~
b_4 = \trans{x}{+y}{-y} \,.\label{eq:b_1_to_b_4}
\end{equation}
The $\polhyperon = \pm y$ ($\polnucleon = \pm y$) denotes a recoil (target) spin quantum number $\pm \frac{1}{2}$ along the $y$ direction. 

\begin{figure*}[tbh]
\includegraphics[width=0.6\textwidth]{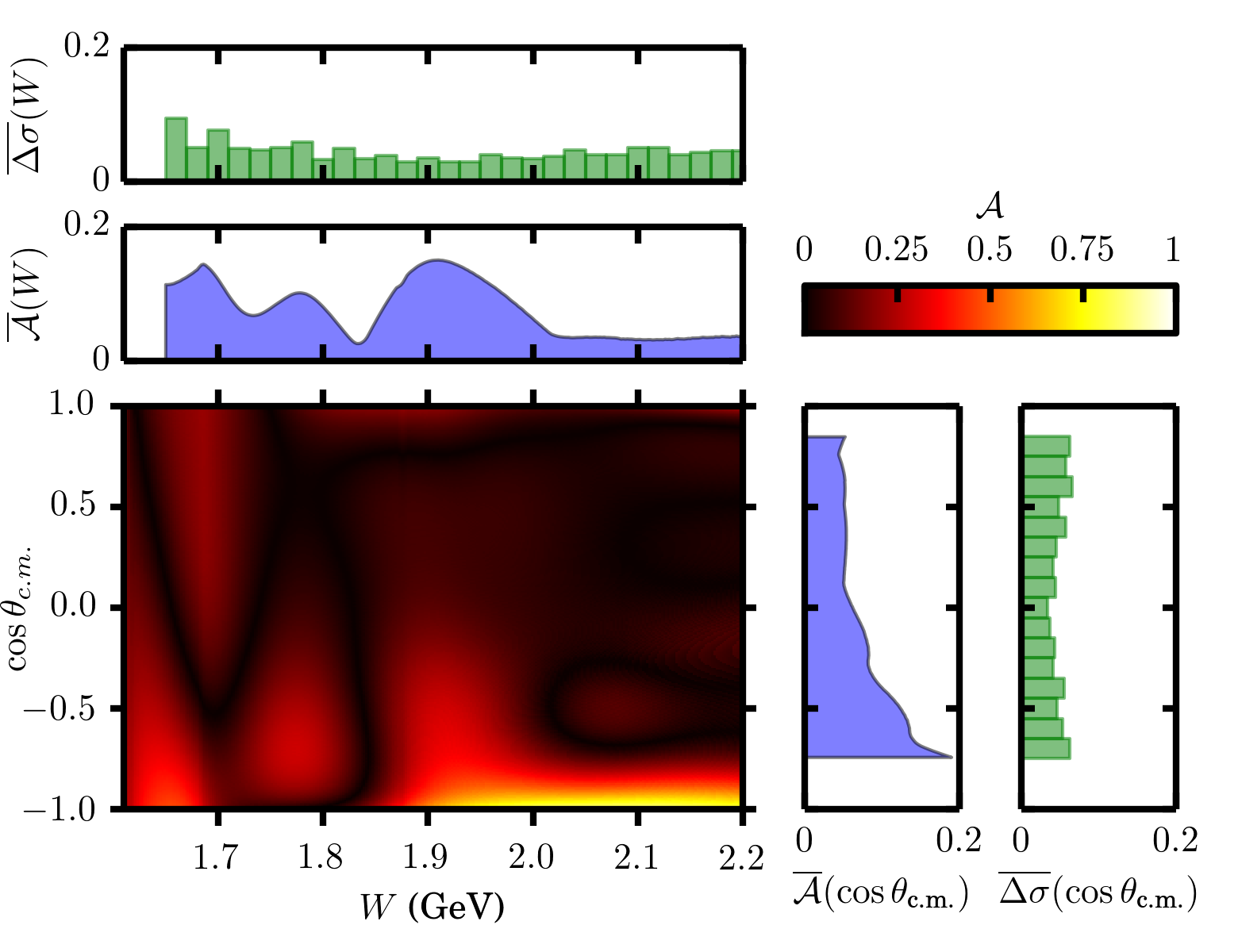}
\caption{
The energy and angular dependence of the $\mathcal{A}$ defined in Eq.\ \eqref{eq:relative_cross_section_distance} between the BoGa and RPR-2011 models for $\gamma p \rightarrow K^{+} \Lambda$.  Also shown are the average $\overline{\mathcal{A}}(\costhcm ) = \frac{1} {b-a} \int _{a} ^{b} d W \mathcal{A}(W,\costhcm)$ 
[a similar formula holds for $\overline{\mathcal{A}}(W)$] in ``realistic kinematics" (RK). Realistic kinematics refers to kinematics accessible with reasonable statistics by existing experimental facilities and is determined by  the ranges $W \geq 1.65$ GeV and $-0.75 \leq \costhcm \leq 0.85$.  The $\overline{\Delta \sigma}(W)$ and $\overline{\Delta \sigma}(\cos \theta _{c.m.})$ are obtained by evaluating the $\gamma p \rightarrow K^{+} \Lambda$  measurements for $\diffcs$. 
Thereby, we have calculated the relative error $ \left(\erroron{\diffcs} \right)/\diffcs$ on an equidistant $(W,\costhcm)$ grid. To compute $\overline{\Delta \sigma}(W)$, for example, we average over the covered $\costhcm$ range at given $W$.       
}
\label{fig:cross_section_differences}
\end{figure*}

In order to quantify the differences between the predictions for the magnitude of the cross sections between the models $A$ and $B$, we introduce the asymmetry
\begin{equation}\label{eq:relative_cross_section_distance}
\mathcal{A}[A,B](W,\costhcm) = \norm{\frac{\diffcs(A) - \diffcs(B)}{\diffcs(A) + \diffcs(B)}} \,.
\end{equation}
In what follows we use the representative Bonn-Gatchina \cite{bogaWebsite,*Anisovich:2014yza,*boga2010} (BoGa) and hybrid Regge-plus-Resonance \cite{rprwebsite,*RPRDeCruzPRL2012,*RPRDeCruzPRC2012,*RPRCorthals2006} (RPR-2011) models for  $\gamma p \rightarrow K^+ \Lambda$ to set the scale of the introduced measure. 
The BoGa  model is a highly sophisticated coupled-channel model. The RPR-2011 model is a hybrid Regge-isobar model for $\gamma p \rightarrow K^{+} \Lambda$ with very low number of parameters. Both models are fitted to a large data set of cross sections, a sizable set of single-polarization observables (mostly $P$) and a limited number of double-polarization observables.  The BoGa and RPR-2011 models parametrize the $\gamma p \rightarrow K^{+} \Lambda$ background very differently at low energies. Figure~\ref{fig:cross_section_differences} shows $\mathcal{A}[A=\text{BoGa},B=\text{RPR-2011}](W,\costhcm)$.  Both models produce comparable cross sections at forward $\thcm$. The results for $\overline{\mathcal{A}}(\cos \theta _{c.m.})$ indicate that the deviations between BoGa and RPR-2011 grow with increasing $\thcm$. This reflects the fact that the description of the background (which requires only a few parameters) in the RPR-2011 model is physically less justified at backward angles \cite{phdDeCruz}.

At extremely backward $\thcm$ and in the threshold region, the measurements typically come with low statistics. Good experimental statistics are obtained for $W \geq 1.65$ GeV and $-0.75 \leq \costhcm \leq 0.85$. In this selected ``realistic kinematics" (RK) the $\mathcal{A}[\text{BoGa},\text{RPR-2011}]$ typically clusters around 0.1-0.2. The experimental equivalent of the asymmetry $\mathcal{A}$ is the relative error $ \left(\erroron{\diffcs} \right)/\diffcs$. The results are included in Figure~\ref{fig:cross_section_differences} and are systematically of the order 0.06 in both $W$ and $\cos \thcm$.  Comparing the $\overline{\mathcal{A}}(\costhcm )$ for (BoGa, RPR-2011) with the experimental figure-of-merit $\overline{\Delta \sigma}(\cos \theta _{c.m.})$, leads us to conclude that the  available experimental information from cross-section measurements in the $\gamma p \rightarrow K^+ \Lambda$ channel is already contained in the BoGa and RPR-2011 models. As a result, further measurements of $\diffcs$ for $\gamma p \rightarrow K^+ \Lambda$ are unlikely to provide information to further discriminate between the assumptions underlying the ``BoGa" and ``RPR-2011" models.


\section{Method for model discrimination in amplitude space}\label{sec:model_model_comparison}
\begin{figure*}[tbh]
\begin{minipage}[t]{0.5\textwidth}
\centering
\includegraphics[width=\textwidth]{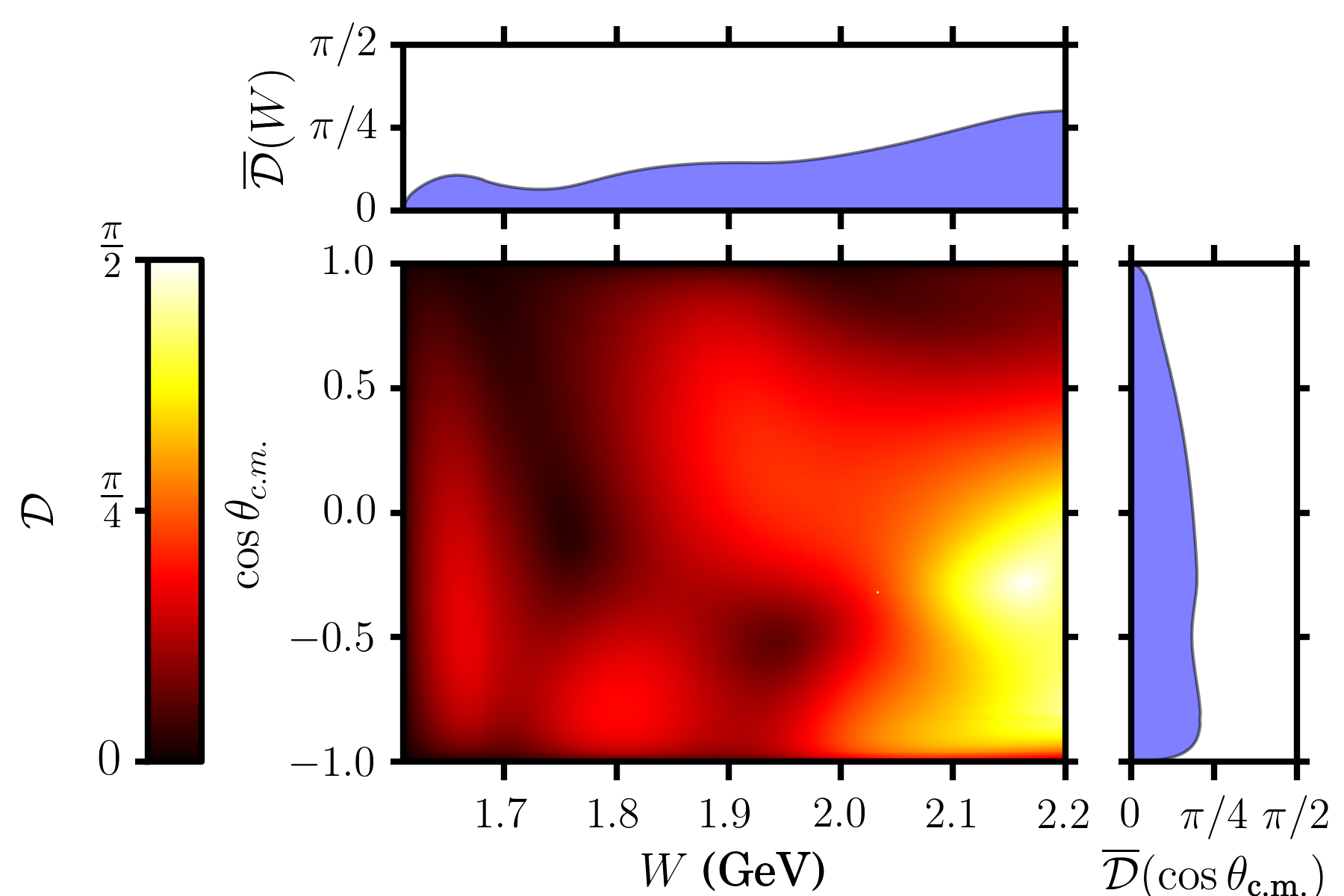}
\end{minipage}%
\begin{minipage}[t]{0.5\textwidth}
\centering
\includegraphics[width=\textwidth]{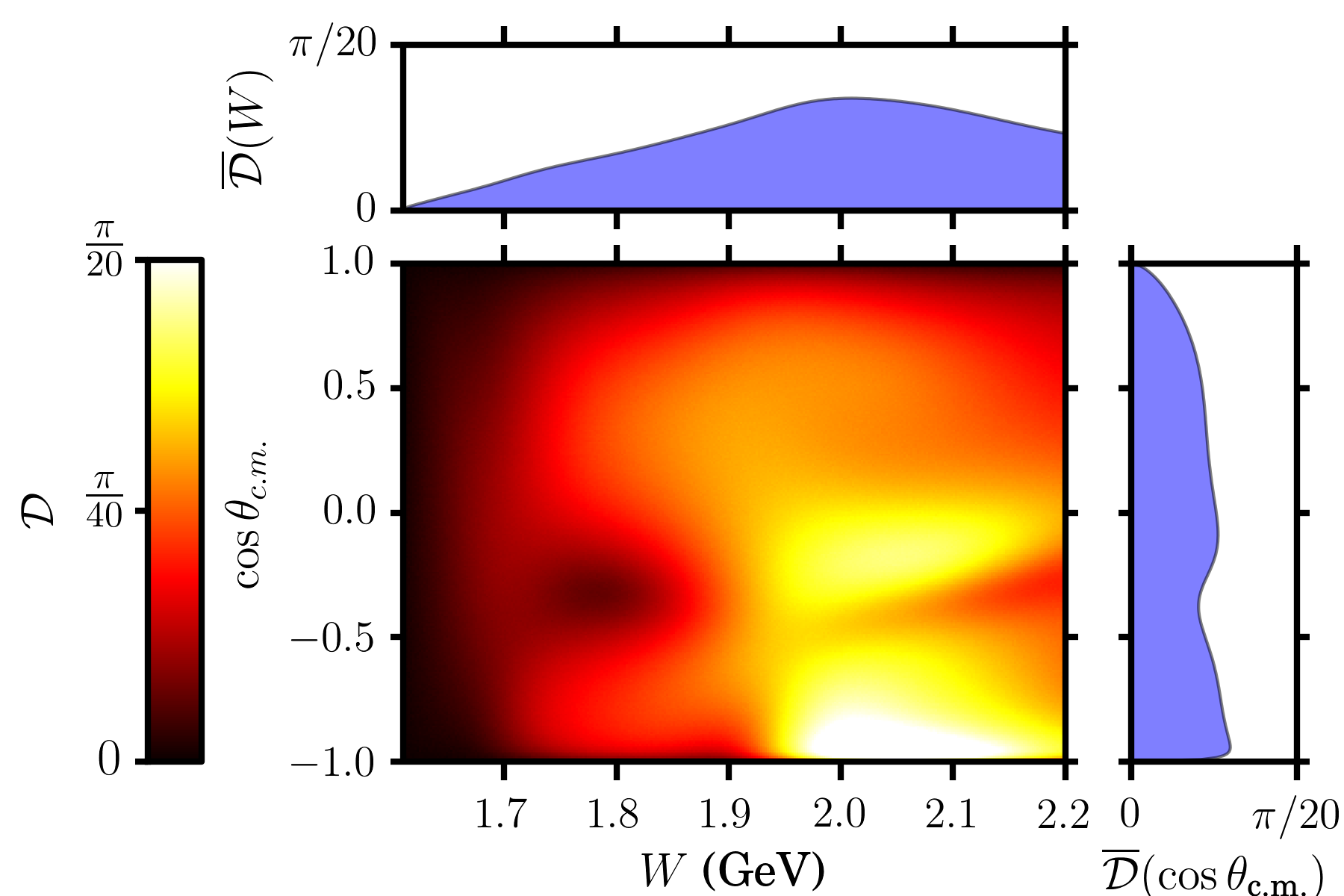}
\end{minipage}
\caption{Energy and angular dependence of the distances $\dist{\text{RPR-2011}}{\text{BoGa}}$ (left) and  $\dist{\text{RPR-2011}}{\text{RPR-2011}^{*}}$ (right) for $\gamma p \rightarrow K^+ \Lambda$. The RPR-2011$^{*}$ differs from RPR-2011 in that the coupling constant for the $D_{13}(1900)$ resonance has been fixed to zero in the fitting process. Note that the color scales are different in both panels. Also shown are the $W$ and $\costhcm$ averaged distances $\overline{\mathcal{D}}(\costhcm)$ and $\overline{\mathcal{D}}(W)$.}
\label{fig:NTA_distances_mapped}
\end{figure*}

To further improve our knowledge of the physics underlying $\gamma N \rightarrow M B$ processes, polarization measurements are key \cite{Strauch:2015zob}. Bilinear relations connect the polarization observables to the amplitudes. Therefore, the potential impact of a polarization measurement is not always clear a priori. 
At given kinematics, a measurement possesses the ability to locally distinguish between two models (or two hypotheses) if its resolving power is smaller than the difference between the two models in amplitude space. Therefore, we introduce a measure to quantify the difference between model $A$ and model $B$ in amplitude space.  
All polarization asymmetries are insensitive to a global scaling factor $Q \equiv \sum_{j=1}^4 \norm{b_j}^2$, and hence, we define the normalized transversity amplitudes (NTA)
\begin{equation}
a_j \equiv \frac{b_j}{\sqrt{Q}} = r_j e^{i\alpha_j} \qquad (j=1,2,3,4)\,.
\end{equation}
All observables are invariant under the global phase transformation $a_j \rightarrow a_j' = a_j e^{i\alpha}$ ($\alpha \in \mathbb{R}$). We define the relative phases $\delta_i^{j} = \alpha_i - \alpha_j$ and introduce the 4D-vector representation of the NTA 
\begin{equation}\label{eq:4d_rep}
\fourv = \left( 
a_{1} \; \; a_{2}  
\; \; a_{3} \; \; a_{4}  
\right)^T \; ,
\end{equation}
which obeys the normalization condition $\fourv ^\dagger \fourv = \sum_{i=1}^4 \norm{a_i}^2 = 1 $. The $3$-sphere in the $\mathbb{C}^4$ representation of Eq.~\eqref{eq:4d_rep} can be mapped onto a geometrically equivalent $7$-sphere in $\mathbb{R}^8$. This analogy provides one with an expression for a
distance in $\mathbb{C}^4$: the opening angle of the position vectors, situated on the surface of the sphere, of two models in amplitude space (measured along the sphere defined by the
normalization condition). For two amplitude
sets $\fourv_A$ and $\fourv_B$ corresponding with the models $A$ and $B$ one can define, 
\begin{equation}\label{eq:distance_measure}
\mathcal{D}[A,B] (W, \costhcm) = \min  \limits_{\alpha_{4} (A)} \left[ \arccos \Re \left( \fourv_A^\dagger  \fourv_B  \right) \right] \,.
\end{equation}
The quantity $\Re \left( \fourv_A^\dagger  \fourv_B  \right) $ depends on the reference phase. As bilinear relations connect the observables to the amplitudes, 
the reference phase is inaccessible, and only the relative phases can be determined. Hence, one can opt to fix $\alpha_4 = 0$ of the $\fourv$'s to infer the phase information from the data, corresponding to the substitution $\alpha_{i} \rightarrow \delta_i^{4} \; (i=1,2,3)$. We wish to provide a distance measure that is independent of the choice of reference phase, so that can it be used both for  comparing models or for comparing a model with data. To this end, in the definition of Eq.~\eqref{eq:distance_measure} we minimize the
distance by varying the reference phase $\alpha_4$ of $\fourv_A$ while keeping all three relative phases $\delta_i^{4}$ in both $\fourv_A$ and $\fourv_B$ fixed.

Figure \ref{fig:NTA_distances_mapped} shows the kinematic dependence of the $\dist{\text{RPR-2011}}{\text{BoGa}}$ for $\gamma p \rightarrow K^+ \Lambda$. The distance between the two models grows as $W$ and $\thcm$ increases. The models differ in their  background parametrization and resonance content.  The differences in the resonance content are mainly visible at backward $\thcm$. Also shown in Fig.~\ref{fig:NTA_distances_mapped} is the kinematic dependence of $\mathcal{D}$ for two versions of the RPR-2011 model. Thereby, we study the kinematic dependence of the distance in amplitude space between the full model and a model variant where the coupling constant of the $D_{13}(1900)$ resonance is forced to be $0$. This allows us to estimate the effect of removing a single resonance on the distance measure of Eq.~\eqref{eq:distance_measure}. 
At forward $\thcm$, where the background contributions dominate, we find that the difference is relatively small compared to backward $\thcm$ where the resonance content dominates. 
The results indicate that confirmation or rejection of the presence of an $s$-channel resonance $R$ from data in a restricted kinematical range, requires experimental resolutions of the order $\mathcal{D} \ll \pi/20$ at backward $\thcm$ and $W \gtrsim M_R$.

\begin{figure}[bth]
\centering
\includegraphics[width=0.5\linewidth]{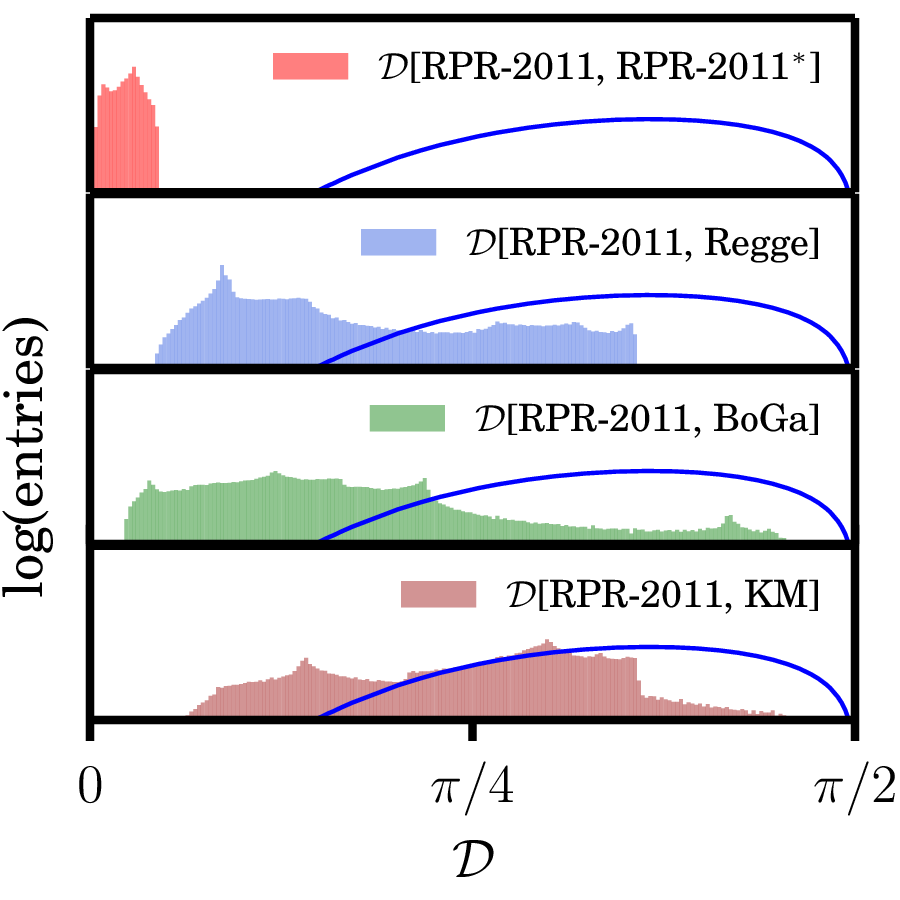}
\caption{%
Logarithm of the entries of $\mathcal{D}[A,B]$ between 4 combinations of two models $\left(\mathcal{M}_A, \mathcal{M}_B \right)$ for $\gamma p \rightarrow K^+ \Lambda$ in the realistic kinematical range (see caption to  Fig.~\ref{fig:cross_section_differences}). The (blue) solid line is the corresponding result for the distances of random samples in NTA space. 
\label{fig:distance_distributions} 
}%
\end{figure}

In order to tell two models $(A,B)$ apart \textit{in amplitude space}, a combined experimental resolution better than the characteristic $\dist{\fourv_A}{\fourv_B}$ is required. In Fig.~\ref{fig:distance_distributions} we show the frequency distributions of $\dist{\fourv_A}{\fourv_B}$, which are derived from  ensembles of values over the range in $W$ and $\cos\theta_{c.m.}$, for four prototypical examples of model combinations $(A,B)$. 
The Kaon-MAID (KM) model \cite{KaonMaidWebsite,*KaonMaidD13} is a prototypical example of an isobar model for $\gamma p \rightarrow K^+ \Lambda$.
The KM model has not been refitted to any data for the past $15$ years, and hence, no double-polarization data was included in the fit. This means that only the moduli of the amplitudes are constrained, while all phases are undetermined by the data. This is reflected in the fact that the values of $\dist{\text{RPR-2011}}{\text{KM}}$ are significant in amplitude space. 
To estimate the required experimental resolution to identify the resonance content, we also show the frequency distribution of $\mathcal{D}$ between the full RPR-2011 model and a pure Regge model that only accounts for the Reggeized background contribution of RPR-2011.  The $\dist{\text{RPR-2011}}{\text{Regge}}$ distribution peaks at considerably larger values than the $\dist{\text{RPR-2011}}{\text{RPR-2011}^{*}}$. This is a reflection of the fact that hunting a particular resonance in a small kinematic interval requires a substantially bigger experimental effort (the results point at a resolution which is at least a factor of five better) than identifying the global effect of all resonances.

\section{Model-independent amplitudes from data}\label{sec:amplitudes_from_data}
\begin{figure*}[tb]
\centering
\begin{minipage}[b]{0.25\linewidth}
\includegraphics[width = \linewidth]{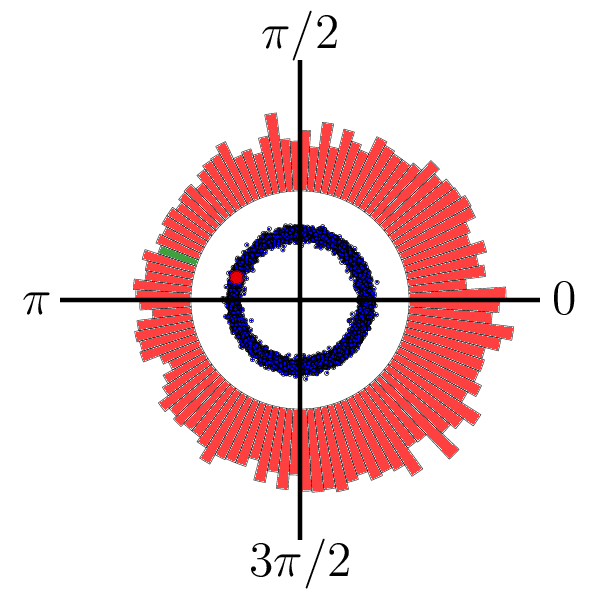}
\end{minipage}%
\begin{minipage}[b]{0.25\linewidth}
\includegraphics[width = \linewidth]{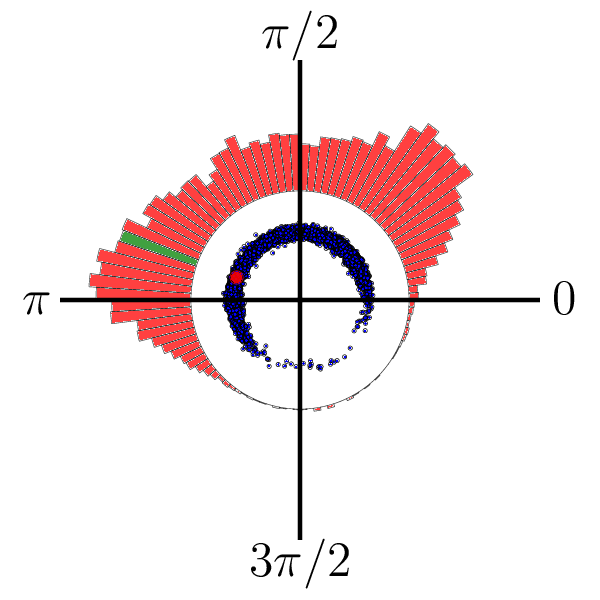}
\end{minipage}%
\begin{minipage}[b]{0.25\linewidth}
\includegraphics[width = \linewidth]{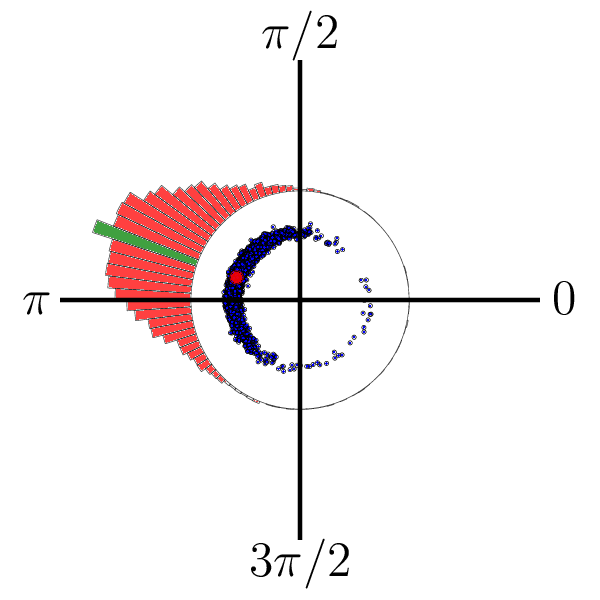}
\end{minipage}%
\begin{minipage}[b]{0.25\linewidth}
\includegraphics[width = \linewidth]{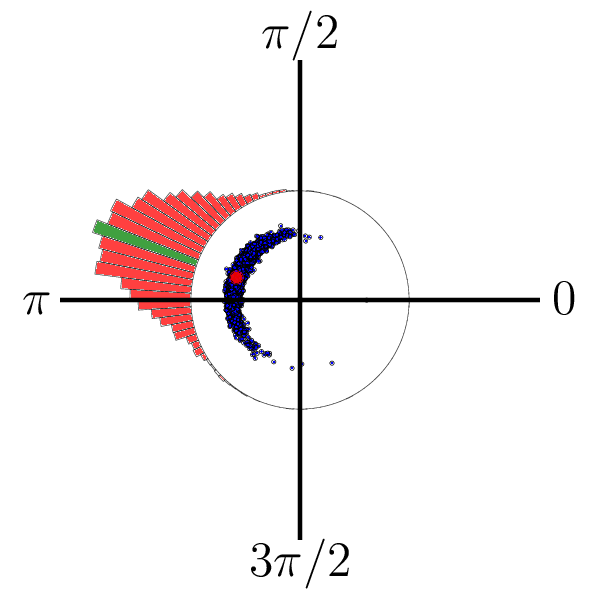}
\end{minipage}
\caption{The distribution of $a_3 e^{-i \alpha_4}= r_3e^{i \delta_3^4}$ at $(W = 1.8 \mbox{ GeV}, \costhcm=-0.1)$ as extracted from ensembles of four different observable sets $\{A_i^{\textrm{exp}} \}$. All observables are distributed as $\mathcal{N}(\mu = A_i^{\textrm{RPR-2011}}, \sigma = 0.1)$. The extracted  $\delta_3^{4}$ are displayed by the red histogram. The blue dots are the bootstrap samples for $r_3e^{i \delta_3^4}$. The red dot and the green bar is the RPR-2011 prediction for $r_3$ and $\delta_3^{4}$. From left to right, we show the $r_3e^{i \delta_3^4}$ extracted  from various combinations of observables: (i) the complete set $\{A_i^{\textrm{exp}} \}_1 = \{\diffcs, \Sigma, T, P, C_x,O_x, E,F\}$ ; (ii) $\{A_i^{\textrm{exp}} \}_2 = \{A_i^{\textrm{exp}} \}_1 + \{C_z, O_z, G\}$ ; (iii) $\{A_i^{\textrm{exp}} \}_3 = \{A_i^{\textrm{exp}} \}_2 + \{H\}$ ; (iv) $\{A_i^{\textrm{exp}} \}_4 = \{A_i^{\textrm{exp}} \}_3 + \{T_x,T_z,L_x,L_z\}$.}
\label{fig:effect_singleKin}
\end{figure*}
This section deals with a fully model-independent extraction of the reaction amplitudes from polarization data. Given a set of measurements, this is in principle achievable using the bilinear relations that connect the observables to the reaction amplitudes. As the data come with finite error bars it is essential to provide realistic estimates of the uncertainties on the extracted amplitudes. We have explored two methods of statistical inference that can quantify the propagated uncertainty on the extracted reaction amplitudes. Firstly, a frequentist approach, using bootstrapping with $\chi^2$ minimization. Secondly, a Bayesian approach whereby we explore the posterior distribution directly. 

Given a set of $N$ measured polarization observables $\{A^{\textrm{exp}}_i \}
\equiv \{A^{\textrm{exp}}_i \pm \erroron{A_i} , i=1,...,N\}$ in a given $(\Delta W, \Delta \costhcm)$ range, a bootstrap method boils down to creating $M$ sets of synthetic data $\{ \{ A_{i}^{(j)} \} \} \equiv \{ \{A_{i}^{(j)} \pm \erroron{A_i}, i=1,...,N \}, j=1,...,M\}$ from $\{A^{\textrm{exp}}_i \}$. Thereby, each observable $A_{i}^{(j)}$ is distributed \footnote{We opt for a normal distribution rather than a beta distribution, since in some published data sets, asymmetry values outside the support $[-1,1]$ range have been reported.} as $\mathcal{N}(\mu = A^{\textrm{exp}}_{i}, \sigma = \erroron{A_i})$. For each of the $M$ sets, the amplitude parameters are inferred by minimizing the cost function
\begin{equation}
\chi^2 \left( \fourv \, ;\, \{A_{i}^{(j)}\}\right) = \sum\limits_{i= 1}^{N} \left(  \frac{A^{\textrm{theo}}_i(\fourv) - A_{i}^{(j)}}{\erroron{A_i}}  \right)^2 \,,
\end{equation}
which results in a set of $M$ amplitude solutions 
\begin{equation}
\fourv^{(j)} = \underset{\fourv}{\textrm{argmin}} \; \; \chi^2 \left( \fourv \, ;\, \{A_{i}^{(j)}\}\right) \qquad j = 1,...,M \; .
\end{equation}
The ensemble $\{ \fourv^{(j)} \}$ can be interpreted as the probability distribution in NTA space of amplitudes that are compatible with the data $\{A^{\textrm{exp}}_i\}$. Each $\chi^2$-inference is a point estimate of the $\fourv$. Therefore, the most likely reaction amplitudes $\fourv$ are those related to the \emph{global} minimum of the $\chi^2$ surface. We search for this minimum with the aid of a genetic algorithm (GA)  followed by a gradient minimizer \cite{root}. This strategy with a combination of a ``rough" and ``high-precision" minimizer algorithm, has already been successfully applied to a precise determination of resonance parameters in Ref.~\cite{Janssen2003,*Ireland:2004kp}. 

Another approach to extracting amplitudes from data is to sample amplitude space and evaluate the log-likelihood function
\begin{align}\label{eq:loglike}
&\loglike (\fourv | \{A^{\textrm{exp}}_{i}\} ) = \sum\limits_{i=1}^N \ln P(A^{\textrm{exp}}_i | \fourv) \\
&= - \frac{N}{2} \ln 2\pi - \sum\limits_{i = 1}^N \ln \erroron{A_i} - \frac{1}{2} \chi^2 \left(\fourv \, ;\, \{A^{\textrm{exp}}_{i}\}\right) \nonumber \,.
\end{align}
Here again we have assumed that all the polarization observables $A^{\textrm{exp}}_i$ are normally distributed. Upon evaluating the Eq.~\eqref{eq:loglike} with the Nested Sampling technique, one also obtains the posterior distribution $P(\fourv | \{ A^{\textrm{exp}}_i \})$. We use the robust MultiNest version of the nested sampling algorithm \cite{feroz2009multinest} in order to obtain posterior samples from  distributions  that may  contain  multiple  modes and pronounced degeneracies in high dimensions. Both the bootstrapping and the Bayesian method described here provide one with a means to understand how uncertainties in the measured experimental observables map onto the probability densities in amplitude space. Obviously the quality of those uncertainties are far superior to for example the Hessian error bars which are often quoted in papers.

As an illustration of the adopted methodology and to convince the reader of the importance of a detailed uncertainty propagation in parameter inference, we illustrate the result of the bootstrap method for the extracted $a_3 e^{-i\alpha_4} = r_3 e^{i\delta_3^{4}}$ at representative kinematics in Fig.~\ref{fig:effect_singleKin}. Thereby we use synthetic data for four combinations of polarization observables. The results indicate that after including realistic error bars for a mathematically complete set as defined by Chiang and Tabakin \cite{ChiangTabakin1997} one is left with a so-called continuous ambiguity with hardly any information about the relative phase of one of the amplitudes. After including information of three more double polarization observables one is left with a multimodal distribution for the phase.  A unimodal posterior is typically reached after including information from $\approx 12$ different polarization observables. Including additional observables now improves the phase resolution via a typical $1/\sqrt{N}$ behavior, where $N$ is the number of observables in the data set.

We quantify the uncertainty of the posterior distribution $P(\fourv | \{ A^{\textrm{exp}}_i \})$ in amplitude space in two ways. First, it is intuitive to regard the posterior as a distribution with a central value and a standard deviation. In Eq.~\eqref{eq:distance_measure}, we introduced a distance measure in amplitude space that quantifies the difference of two models at $(W,\costhcm)$. Obviously, in order to tell the different models apart, one should aim at carrying out experiments with a resolving power better than those representative values. In the absence of any data, the NTA are uniformly distributed over the surface of a unit 7-sphere. In what follows we refer to this distribution as the prior $\pi(\fourv)$. We work out the evolution of the resolution in amplitude space reached after combining data from several single- and double-polarization experiments. Using Eq.~\eqref{eq:distance_measure}, the dispersion of the ensemble of NTAs can be readily computed from 
\begin{equation}\label{eq:sigma}
\erroron{\fourv} = \sqrt{\left\langle \dist{\fourv_0}{\fourv}^2 \right\rangle_{P(\fourv|\{ A^{\textrm{exp}}_i \})}} \, ,
\end{equation}
where $\fourv_0$ is the central amplitude vector. It can be shown that $0 \leq \mathcal{D} \leq \pi/2$. For a given  $\fourv_0$, one finds $\erroron{\fourv} \approx 1.08$ for $\fourv$ distributed according to the prior $\pi(\fourv)$. 
Note that expression \eqref{eq:distance_measure} and the $\erroron{\fourv}$ are invariant under any unitary transformation of the amplitudes. Hence, Eq.~\eqref{eq:sigma} yields results which are identical for all amplitudes bases which are connected through unitary transformations. For example, the NTA and the normalized helicity amplitudes (NHA) result in identical values for $\erroron{\fourv}$. 

\begin{figure*}[tbh]
\includegraphics[width=0.7\textwidth]{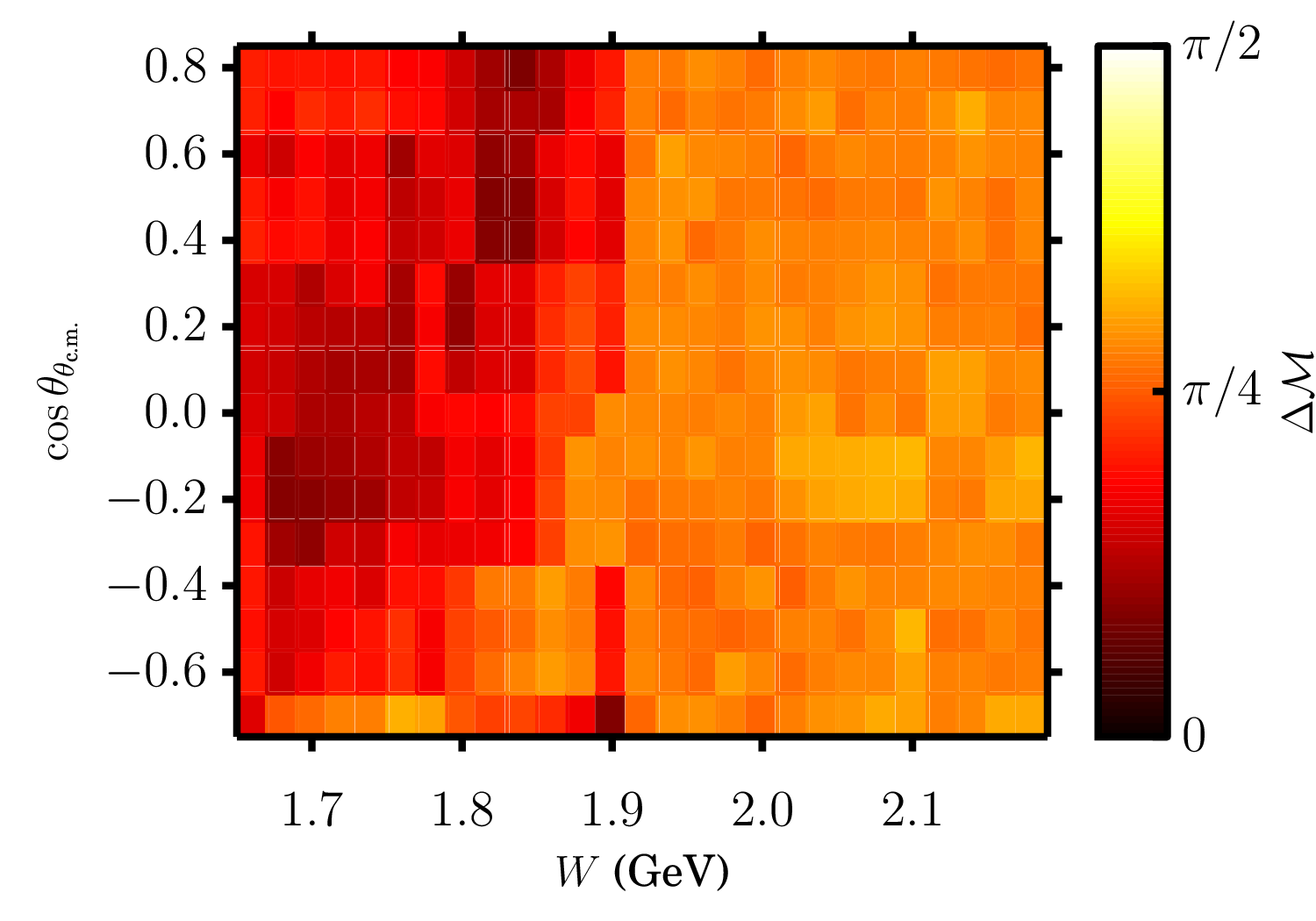}
\caption{The kinematic dependence of the computed $\erroron{\fourv}$ given the published  $\gamma p \rightarrow K^+ \Lambda $ polarization data.
The grid is determined by $(\Delta W=20~\text{MeV}, \Delta \costhcm =0.1)$.}
\label{fig:extraction_maps}
\end{figure*}

Figure~\ref{fig:extraction_maps} illustrates the current status of $\erroron{\fourv}$ given the published polarization data for $\gamma p \rightarrow K^+ \Lambda$.  Hereby, we use the available CLAS  $\{ P, C_{x',z'} \}$ data \cite{Bradford2006, McCracken2010, McNabb2004, Bradford2007} and GRAAL  $\{ \Sigma, T, P, O_{x,z} \}$ data \cite{Lleres2007,Lleres2009}. For $W>1.9$~GeV, one obtains larger $\erroron{\fourv}$ values, which is a reflection of the fact that the GRAAL data extends from threshold to $W \lesssim 1.9$~GeV. Inclusion of new CLAS data \cite{paterson_photoproduction_2016}, which covers a wider energy range, lowers $\erroron{\fourv}$ for $W \geq 1.9$~GeV to values that are comparable to those obtained for $W<1.9$~GeV in Fig.~\ref{fig:extraction_maps}. Interestingly, inclusion of the new CLAS data does not significantly diminish the $\erroron{\fourv}$ values for $W <1.9$~GeV. This is primarily due to the fact that $\{ \Sigma, T, P, C_{x,z}, O_{x,z} \}$ is not a mathematically complete set of observables.

The standard deviation in Eq.~\eqref{eq:sigma} is useful to connect the resolving power of  experiments to the distance between models \eqref{eq:distance_measure}.
The second method to quantify the uncertainty of the posterior distribution $P(\fourv | \{ A^{\textrm{exp}}_i \})$ 
does not require a central value $\fourv_0$. 
In Ref.~\cite{Ireland} it was pointed out that information entropy is a convenient way of quantifying the extent to which one reaches a status of practical completeness given a set of measurements.  The information entropy $H(P)$ of the posterior $P(\fourv|\{ A^{\textrm{exp}}_i \})$ is defined as
\begin{equation}\label{eq:entropy}
H(P) = \int  P(\fourv|\{ A^{\textrm{exp}}_i \}) \log_2 P(\fourv|\{ A^{\textrm{exp}}_i \}) \, \mathrm{d} \fourv \,.
\end{equation} 
We calculate the information gained through measurements relative to the prior distribution $\pi(\fourv)$ (which reflects the situation of no measured polarization observables) 
\begin{equation}\label{information_gain}
I(\pi, P) = H(\pi) - H(P) \; .
\end{equation}
A large information gain indicates that a set of measurements accomplishes an exclusion of significant parts of the domain of possible solutions in amplitude space. Since we choose to use base-2 $\log$s, information is quantified in bits. One bit of information is equivalent to decisive information on a boolean decision. For example, assume a set of measurements which leaves a discrete ambiguity, corresponding to two identical, but non-overlapping peaks in amplitude space. An additional measurement of which the only effect is that it completely excludes one of the two solutions, corresponds to an information gain of exactly one bit.

\begin{figure*}[bth]
\centering
\begin{minipage}{0.3\textwidth}
\includegraphics[width=\textwidth]{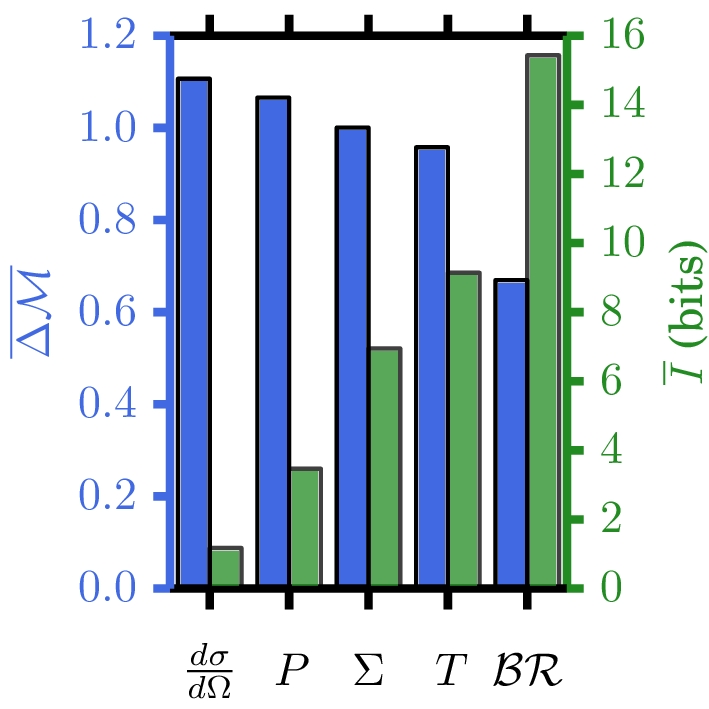}
\end{minipage}%
~
\begin{minipage}{0.63\textwidth}
\includegraphics[width=\textwidth]{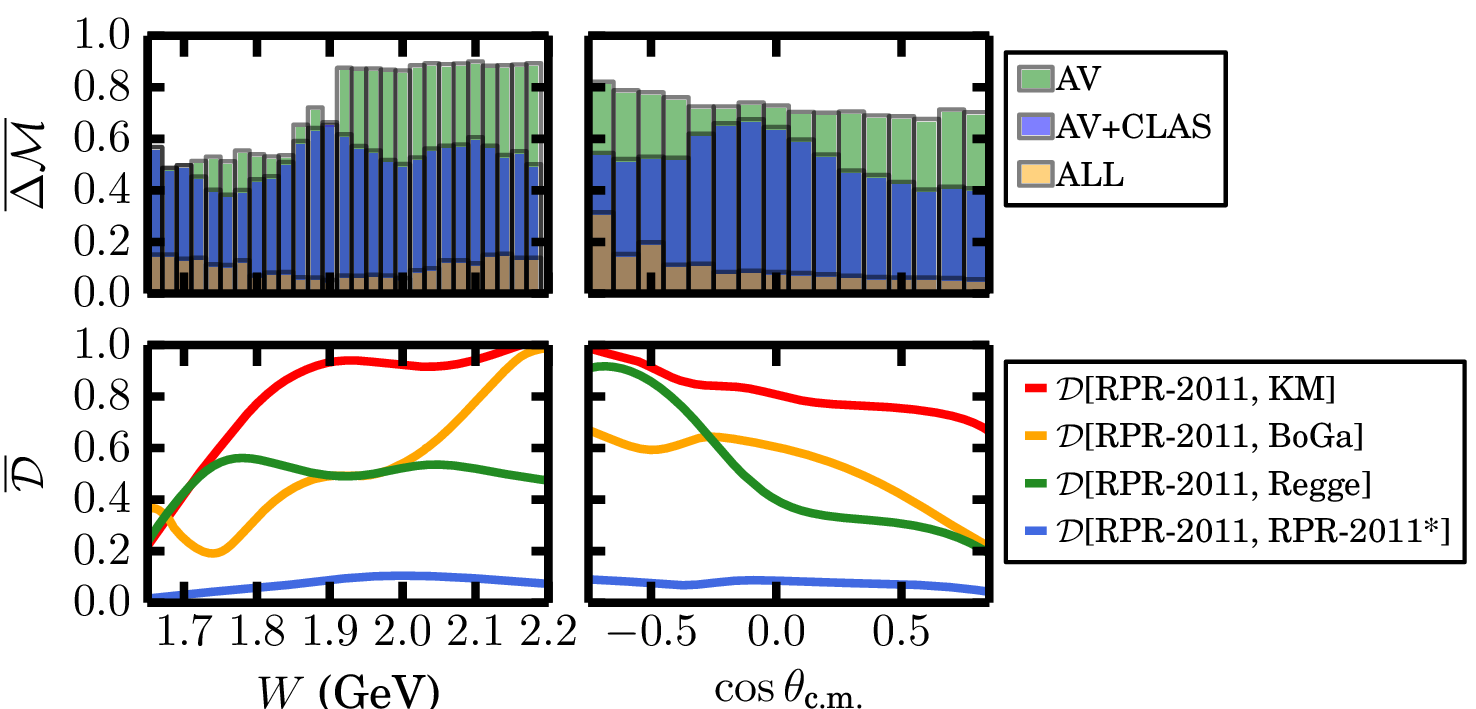}
\end{minipage}
\caption{(Left) The $\costhcm$-averaged amplitude uncertainty $\erroron{\fourv}$ and information gain $I$ for the $\gamma p \rightarrow K^+ \Lambda$ reaction at $1.89 \leq W \leq 1.91$~GeV. From left to right the number of observables included cumulatively grows as indicated on the axis. (Right) The $W$- and $\costhcm$-averaged model distances $\mathcal{D}$ and the anticipated $\erroron{\fourv}$. The $\mathcal{D}$  
are evaluated for four different model combinations. The $\erroron{\fourv}$ is shown for amplitude extraction with the published CLAS and GRAAL data (``AV"). The ``AV+CLAS" results for $\erroron{\fourv}$ include also the as yet unpublished CLAS polarization data.  The ``ALL" results for $\erroron{\fourv}$ are obtained with synthetic data (with realistic error bars) for all  $15$ possible polarization observables. For example, to generate $\overline{\erroron{\fourv}}(W)$ for the observable set ``AV'', the results in Fig.~\ref{fig:extraction_maps} have been averaged over the considered $\costhcm$ range.}
\label{fig:extraction_results}
\end{figure*}

The left panel in Figure \ref{fig:extraction_results} shows the result of a Bayesian inference of the reaction amplitudes for a number of observable sets. None of the observable sets constitute a complete set. Hence, there is an upper value by which the information gain is limited. Using the available data, one can at best  determine the moduli $r_{i = 1,..,4}$ and two relative phases $\left( \delta_1^{4}, \delta_2^{3} \right)$. It was estimated in \cite{Ireland} that approximately $21$ bits of information gain are required to form a well-defined unimodal distribution. 

The comparison of the expected data resolution to the benchmark model distances are also depicted in Fig.~\ref{fig:extraction_results}. For the available data set and $W<1.9$ GeV, we obtain $\erroron{\fourv} \sim 0.5$, which indicates that one can locally resolve between RPR-2011 and the Kaon-MAID models. Also, the uncertainty on the available data is low enough to locally distinguish RPR-2011 and the pure Regge model. Hence, even in a limited kinematical region (``locally"), we can at least say there is evidence of $s$-channel resonances in the current data.
For $W > 1.9$ GeV, the distance $\mathcal{D}[\textrm{RPR-2011}, \textrm{Regge}]$ is less than the spread $\erroron{\fourv}$ of the available data, but similar to the resolution provided by the new CLAS data in addition to available data. Therefore we expect that the new data should be able to tell us more about the existence of resonances in the RPR model above $W = 1.9$ GeV, while the effect below this energy is relatively modest. Since background contributions dominate at forward angles, $\mathcal{D}[\textrm{RPR-2011}, \textrm{Regge}]$ and $\mathcal{D}[\textrm{BoGa}, \textrm{Regge}]$ fall from backward to forward $\thcm$, therefore it is apparent that measurements at backward angles contain more information about the presence of $s$-channel resonances. The existence of specific resonances has a very small effect in amplitude space at a single $(s,t)$ point.
It is also observed that measurements of all observables are required to resolve the relatively small distance between models differing by one resonance.

Summarizing, we have investigated methods to quantify the distance between models in amplitude space. This distance measure can also be used to estimate in a model-independent way how well a given set of combinations of polarization measurements will succeed in constraining the underlying reaction amplitudes. 


\begin{acknowledgments}
This work was supported by the Research Foundation Flanders (FWO-Flanders) and the United Kingdom's Science and Technology Facilities Council (STFC) from grant number ST/L005719/1. J.~Nys was supported as an 'FWO-aspirant'. The computational resources (Stevin Supercomputer
Infrastructure) and services used in this work were provided by Ghent University, the Hercules Foundation 
and the Flemish Government.
\end{acknowledgments}

\FloatBarrier
\newpage

\bibliographystyle{biblio-physrev}
\bibliography{biblio.bib}

\end{document}